\documentclass{aastex61}

\newcommand\aastex{AAS\TeX}

\received{}
\revised{}
\accepted{}
\submitjournal{ApJ}

\shorttitle{\aastex\ Cosmological Nucleosynthesis}
\shortauthors{A. Maeder}

\begin{document}

\title{Evolution of the early Universe in the scale invariant theory}


\correspondingauthor{Andre Maeder}
\email{andre.maeder@unige.ch}

\author[0000-0001-8744-0444]{Andre Maeder} 
\affiliation{Geneva Observatory \\
chemin des Maillettes, 51 \\
CH-1290 Sauverny, Switzerland}

\begin{abstract}
Analytical solutions are obtained for the early cosmological phases in the scale invariant models with curvature $k=0$.
They complete the analytical solutions already found in the matter--dominated era by \citet{Jesus18}. The physical properties 
in the radiative era are derived from the conservation laws and compared to those of current standard models. 
The critical runs of the temperature  $T$(MeV) 
and of the expansion rate $H$
of the scale invariant models   with low densities, {\it{e.g.}} $\Omega_{\mathrm{m}}=0.04$, 
 are quite  similar at the time of nucleosynthesis to those of standard models with  $\Omega_{\mathrm{m}}=0.30$,
leading to the same freezing number ratio of neutrons to protons.
 These results are consistent with the fact that the scale invariant models
appear to not require the presence of dark matter.

\end{abstract}


\section{Introduction} \label{sec:intro}

It is often ignored that  when the cosmological constant $\Lambda$ is assumed to be equal to zero, the equations of General Relativity 
are scale invariant, which is also a property present in Maxwell equations of electrodynamics. However,
 for a non-zero $\Lambda$,  the field equations of the gravitation no longer show the property of scale invariance.
 A fact, which as discussed by \citet{Bondi90},  was one of the reasons of Einstein's disenchantment for the cosmological constant.
 It is thus of interest to examine at what conditions the scale invariant properties of General Relativity may be restored,
 since current cosmological observations appear to support a  positive cosmological constant.
 
A theoretical framework has been developed by \citet{Dirac73} and \citet{Canu77}, in
the so-called co-tensorial calculus based on the Integrable Weyl's  geometry. It offers a consistent basis to account for 
the properties of scale invariance of gravitation to a scale factor $\lambda$, 
as also illustrated by several properties studied by \citet{BouvierM78}.
Scale invariant derivatives, modified affine connexions, modified Ricci tensor and curvatures can be obtained  leading to a general scale
invariant field equation. Dirac and  Canuto et al. have  expressed
 an action principle in the scale invariant framework, with   a matter Lagrangian, as a coscalar  of power -4 (varying like $\lambda^{-4}$).
 By considering the variations of this  action, they also obtain the generalization of the Einstein field equation. 
 This is  Equation (7) in \citet{Maeder17a}  from  which the scale invariant cosmological equations  derived. 
 
 In the integrable  Weyl’s space,  the scale factor remains undetermined without any other constraints
 as shown by Dirac and Canuto et al. Thus, these authors were fixing the scale factor by some external 
 considerations based on the so-called  Large Number Hypothesis, an hypothesis often disputed \citep{Carter79}. Thus, it seems 
 appropriate  to explore other conditions for setting the gauge.
 The proposition was thus made to fix the gauge factor by simply assuming  the  scale invariance of the empty space \citep{Maeder17a}. 
 This means that the properties of the empty space should not change under a contraction or dilatation of the space-time.
 Indeed, we note, as shown by  \citet{Carr92}, that the current
 equation of the vacuum $P_{\mathrm{vac}}= - \varrho_{\mathrm{vac}} \, c^2$  already implies that    
 $\varrho_{\mathrm{vac}}$ should remain constant ''if a volume of
 vacuum is adiabatically compressed or expanded''. On this basis, the cosmological equations derived by \citet{Canu77} were greatly
 simplified \citep{Maeder17a}.
 A number of cosmological tests were performed, with positive results. Interestingly enough, these equations were
 then shown to have rather simple analytical solutions  \citep{Jesus18} for models of a matter dominated Universe with a zero curvature.
 
 In order to express the  motions of free particles, a geodesic equation was obtained  \citep{BouvierM78}
  from a minimum action in the integrable Weyl’s space. In the weak field approximation,
  the geodesic equation leads to a  modification of the Newton equation \citep{MBouvier79}: it then contains
  a (currently very small) additional acceleration term proportional to the velocity of the particles. This equation was applied
  to study the internal motions of clusters of galaxies, the flat rotation curves of spiral galaxies and  the age increase
  of the ''vertical'' velocity dispersion of stars in galaxies \citep{Maeder17c}. 
  The interesting result was that the  observational properties of these various systems
  could be accounted without requiring to  the current hypothesis of dark matter, and the same for the radial acceleration relation (RAR)
  of galaxies \citep{Maeder18}.
  The growth of the density fluctuations in the early Universe was also studied by
  \citet{MGueorguiev18}. The usual theory is that at recombination the baryons settle in the potential wells formed by dark matter 
  previously assembly during the radiative era. In the scale invariant framework, the growth of the density fluctuation, predicted from
  the corresponding Euler, Poisson and continuity equations, is fast enough
  so that dark matter is not needed to account for the formation of galaxies.
  
  The object of the present work is to search for the solutions  in the radiative era of the universe in the scale invariant context.
    In Section 2, we derive the analytical solutions for cosmological
  models dominated by relativistic particles, and examine the values at the equilibrium of matter and radiation. In Section 3, 
  the physical properties in the radiative era are studied. The conditions at the time of the cosmological nucleosynthesis 
  are compared to the results of standard models.
    \\

\section{The cosmological solutions in the radiative era}

We study   the  physical conditions in the radiative era, in particular when the cosmological nucleosynthesis occurs.
At this stage,  the age of the Universe, of 1 second and more, is already $10^{43}$ times larger than the Planck time
and  thus the Universe dynamics is no longer   dominated by quantum effects. Thus, if it occurs that scale invariance 
applies in the present day Universe, it should likely also be the case at the time of nucleosynthesis.
For the matter era, analytical 
solutions were found by \citet{Jesus18} in the case of a flat Universe model with $k=0$. Now, we also need the
solutions of the cosmological models in the radiative era.

 \subsection{The equations and their solutions} \label{theory1}
 
 The hypothesis of scale invariance assumes that the equations should be invariant to a transformation of the form 
  $ds'= \lambda(x^{\mu}) ds$ \citep{Weyl23,Dirac73,Canu77}. The term $\lambda$ is the scale factor expressing a relation of the form $ds'= \lambda(x^{\mu}) ds$, 
where $ds'$ is the line element in general relativity and $ds$ the line element in the supposed more general scale invariant space.
For reasons of homogeneity and isotropy the scale factor depends only on time $t$,
and the condition of the invariance of the empty space
imposes  that $\lambda \sim 1/t$ \citep{Maeder17a}. Interestingly enough, this form is also one of those considered by \citet{Dirac73}.
 The  equations of the scale invariant cosmology  under the above mentioned hypothesis are 
 \begin{equation}
\frac{8 \, \pi G \varrho }{3} = \frac{k}{R^2}+\frac{\dot{R}^2}{R^2}+ 2 \,\frac{\dot{R} \dot{\lambda}}{R \lambda} \, ,
\label{E1}
\end{equation} 
\begin{equation}
-8 \, \pi G p  = \frac{k}{R^2}+ 2 \frac{\ddot{R}}{R}+\frac{\dot{R^2}}{R^2}
+ 4 \frac{\dot{R} \dot{\lambda}}{R \lambda}  \, .
\label{E2}
\end{equation}
\noindent
The combination of these two equations  leads to
\begin{equation}
-\frac{4 \, \pi G}{3} \, (3p +\varrho)  =  \frac{\ddot{R}}{R} + \frac{\dot{R} \dot{\lambda}}{R \lambda}  \, .
\label{E3}
\end{equation}
\noindent
The various terms have their usual significations, with dots representing  time derivatives. These equations only differ from the 
usual ones by the presence in each case  of an additional term containing the product $\frac{\dot{R} \dot{\lambda}}{R \lambda} $.
We see that if $\lambda(t)$ is a constant, 
the usual equations of cosmologies are recovered. 
This means that, at any fixed time, the effects that do not depend on time evolution are just those 
predicted by GR.

For the flat space with $k=0$ in the matter dominated era,
 this equation has an analytical solution of the following form  \citep{Jesus18}, 
\begin{equation}
R(t)  \,= \, \left(\frac{t^3- \Omega_{\mathrm{m}}} {1- \Omega_{\mathrm{m}}} \right)^{2/3} \,,
\label{jesus}
\end{equation}
\noindent
where $\Omega_{\mathrm{m}} = \varrho_{\mathrm{m,0}}/\varrho_{\mathrm{c,0}}$ 
with  $\varrho_{\mathrm{c,0}}= 3 \, H^2_0/(8 \, \pi G)$, the standard expression of  the critical density at present.
In the above equation, it is assumed that at the
present  time $t_0=1$, one has $R_0 =1$.
Interestingly enough, it is also possible  in this context  to have  for a flat space different values of $\Omega_{\mathrm{m}}$.
The expression  of the Hubble expansion  term $H(t)$ resulting from (\ref{jesus}) is 
\begin{equation}
H(t) \, = \, \frac{2 \, t^2}{t^3- \Omega_{\mathrm{m}}} \,, \quad \mathrm{so \; that}\quad H_0\, = \, \frac{2}{1- \Omega_{\mathrm{m}}} \,,
\label{Ht}
\end{equation}
\noindent
 expressed here in the above scales of time and space.
  In Sect. \ref{physrad}, these quantities will be used with their appropriate physical units. 
 We see from these two expressions that there is no solutions for $\Omega_{\mathrm{m}} \geq 1$, 
 since the presence of  a significant matter content in the Universe is killing scale invariance \citep{Maeder17a}.\\

The radiative era of the Universe is dominated by relativisitic particles. The density  is $\varrho_{\mathrm{rel}}$,
\begin{equation}
\varrho_{\mathrm{rel}} \, = \, K_0 \,\varrho_{\gamma}\, , \quad \mathrm{with} \; \; 
 K_0\,= 1+ \frac{7}{8} \left(\frac{4}{11} \right)^{4/3}\,N_{\nu} \, .
\end{equation}
\noindent
There, $\varrho_{\gamma}$ is the  density of photons. The term $K_0$ accounts for the neutrino contribution to the density
of relativistic particles \citep{Coles02}, where $N_{\nu}$ is the number of neutrino types. For a number $N_{\nu}=3$, the value is
$K_0=1.6813$.  The appropriate conservation law 
for relativistic particles has to be accounted for. According to \citet{Maeder17a}, 
it is $\varrho_{\mathrm{rel}} \, R^4 \lambda^2= const.$, thus 
Equation (\ref{E1}) writes, 
\begin{equation} 
  \lambda^2 \dot{R}^2 R^2 + 2 \dot{R} \,R^3 \dot{\lambda} \,\lambda  \,= -  C_{\mathrm{rel}} \, ,
\end{equation}
with
\begin{equation}
C_{\mathrm{rel}}\, = \, \frac{8 \, \pi G \varrho_{\mathrm{rel}} R^4 \lambda^2}{3} \, .
\label{crel}
\end{equation}
The differential equation to be solved for $R(t)$ is thus,
\begin{equation}
\dot{R}^2 R^2 \, t  -2 \dot{R}R^3 - C_{\mathrm{rel}}\, t^3= \, 0 \,.
\label{E4}
\end{equation}
We  now examine whether 
 the above differential equation also possesses analytical solutions. We try to develop $R(t)$ as a product,
 \begin{equation}
 R(t) = t \, v(t)\, , \quad \mathrm{thus} \quad \dot{R}= v + t \, \dot{v} \,.
 \label{vt}
 \end{equation}
 \noindent
 Thus, Equation (\ref{E4}) leads to
 \begin{equation}
 \dot{v}\, = \,  \pm{ \sqrt \frac{v^4+C_{\mathrm{rel}}}{v^2 t^2}} \, , \quad \mathrm{and} \quad
\frac{v dv}{\sqrt{v^4+C_{\mathrm{rel}}}} \, = \pm \frac{dt}{t} \, .
\end{equation}
The equation is separable and from  \citet{Bronstein74} we obtain as an integral
 \begin{equation}
\ln \, \left(\sqrt{v^4+C_{\mathrm{rel}}} + v^2 \right) \, = \pm  2\, \ln t + const. 
 \end{equation}
 Considering solutions of $R(t)$ growing with time, we take  the sign ''+'' in the above equation
 and obtain
 \begin{equation}
 \sqrt{v^4+C_{\mathrm{rel}}} + v^2 \, = c_2 \, t^2 \, ,
 \label{c2}
 \end{equation}
 \noindent 
 where $c_2$ is a constant, which will be fixed below from the relevant boundary conditions.
 The present solution does not apply at time $t_0$, but
 only in the radiative era, thus  it is not possible to directly express the constants $ C_{\mathrm{rel}}$ and $c_2$
 as functions of  cosmological parameters (like $H_0$) at the present time. We get after some simple manipulations
 \begin{equation}
 v \, = \, \left(\frac{c_2^2t^4 - C_{\mathrm{rel}}}{2 \,  c_2 t^2} \right)^{1/2}
 \label{v}
 \end{equation}
 \noindent
 and for the expansion factor
 \begin{equation}
 R(t) \, = \,t \,  \left(\frac{c_2^2t^4 - C_{\mathrm{rel}}}{2 \,  c_2 t^2} \right)^{1/2} \, =
  \, \frac{1}{\sqrt{2}} \left(c_2 \, t^4 -\frac{C_{\mathrm{rel}}}{c_2} \right)^{1/2} \,.
 \label{R}
 \end{equation}
 \noindent
 The derivative of the expansion factor is
 \begin{equation}
 \dot{R} \, = \, \left(\frac{c_2^2 \, t^4 - C_{\mathrm{rel}}}{2 \,  c_2 t^2} \right)^{1/2}+\frac{t}{2}\,
  \frac{c^2_2 \,t^4+C_{\mathrm{rel}}}{c_2 \, t^3}  \,\left(\frac{2 \,c_2\,t^2}{c^2_2\,t^4- C_{\mathrm{rel}}} \right)^{1/2} \, ,
 \end{equation}
 \noindent
 so that the Hubble expansion rate at a given time in the radiative era is,
 \begin{equation}
 H \, = \, \frac{\dot{R}}{R} \, = \, \frac{1}{t}+\frac{1}{2} \frac{t^2}{R^2} \,
 \frac{c^2_2\, t^4+C_{\mathrm{rel}}}{c_2 \,t^3}\, .
 \label{Hrad}
 \end{equation}
 There is thus an analytical solution of the scale invariant equation in the radiative era, which may greatly simplify all
 studies. The solution depends  on the constants $C_{\mathrm{rel}}$ and $c_2$, 
 which we shall relate below to the  densities of matter and of relativistic particles, respectively $\varrho_{\mathrm{m}}$  
 and $\varrho_{\mathrm{rel}}$ at the time when these two densities become equal.

\subsection{Values of the constants and properties at the density  equilibrium}   \label{equil}

In the radiation era, according to Equation (\ref{crel}) which expresses  the appropriate conservation law, 
the density of relativistic particles at a given time $t$ is,
\begin{equation}
\varrho_{\mathrm{rel}} \, = \, \frac{3 \, C_{\mathrm{rel}} }{8 \, \pi G R^4 \lambda^2} \, .
\end{equation}
In the matter era, the corresponding expression obtained from the matter conservation law is \citep{Maeder17a},
\begin{equation}
\varrho_{\mathrm{m}} \, = \, \frac{3 \, C_{\mathrm{m}} }{8 \, \pi G R^3 \lambda} \, .
\end{equation}
\noindent
At the time $t_{\mathrm{eq}}$ when both densities become equal, we have the following relation between the constants,
\begin{equation}
C_{\mathrm{rel}}  \,=\, C_{\mathrm{m}} \, \frac{R_{\mathrm{eq}}}{t_{\mathrm{eq}}} =
 \, \frac{4 \, \Omega_{\mathrm{m}}}{(1-\Omega_{\mathrm{m}})^2} \, \frac{R_{\mathrm{eq}}}{t_{\mathrm{eq}}} \, ,
\label{CC}
\end{equation}
\noindent
where $R_{\mathrm{eq}}$ and $t_{\mathrm{eq}}$ are the values at equilibrium and the constant 
$C_{\mathrm{m}}=  \frac{4 \, \Omega_{\mathrm{m}}}{(1-\Omega_{\mathrm{m}})^2}$
 in the matter dominated era,  as given by Equation (66) in \citet{Maeder17a}.
 The term $\Omega_{\mathrm{m}}$ is the 
 usual parameter of  matter density as discussed in Section (\ref{theory1}).  
The constant $c_2$ is fixed by expression (\ref{c2}) applied  to the time of equilibrium,
\begin{equation}
c_2= \frac{v^2_{\mathrm{eq}}+\sqrt{v^4_{\mathrm{eq}}+C_{\mathrm{rel}}}}{t^2_{\mathrm{eq}}} \, ,
\quad  \mathrm{with} \quad  v_{\mathrm{eq}} \, = \, \frac{R_{\mathrm{eq}}}{t_{\mathrm{eq}}} \,.
\label{c2eq}
\end{equation}
\noindent 
The expression  of $ v_{\mathrm{eq}}$ in terms of basic quantities is given below in Equation (\ref{Rt}).
The values of $R_{\mathrm{eq}}$ and $t_{\mathrm{eq}}$ are fixed thanks to the conservation laws. In the matter and radiative eras,
we respectively have at a time $t$,
\begin{equation}
\varrho_{\mathrm{m }}\,R^3 \lambda \, = \,\Omega_{\mathrm{m}} \, \varrho_{\mathrm{c,0}} \, ,  \quad \mathrm{and}
\quad  \varrho_{\mathrm{rel}}\, R^4 \lambda^2 \,  =  \, \varrho_{\mathrm{rel,0}} \, ,
\label{conserv}
\end{equation}
\noindent
where the index $_0$ indicates values at the present time $t_0$.
The equality of the two densities defines the equilibrium point, we get
\begin{equation}
 R_{\mathrm{eq}} \, \lambda_{\mathrm{eq}}\, =\, \frac{R_{\mathrm{eq}}}{t_{\mathrm{eq}}} \,\left(= \,v_{\mathrm{eq}}\right) \, = 
\, \frac{K_0 \, \varrho_{\mathrm{\gamma,0}}}{\Omega_{\mathrm{m}} \,  \varrho_{\mathrm{c,0}}}\, .
\label{Rt}
\end{equation}
\noindent
The numerical value of $v_{\mathrm{eq}}$ is given by the above expression, then   $t_{\mathrm{eq}}$  is obtained 
by using the expression for $R(t)$ in the matter dominated era, as given
by Equation (\ref{jesus}). We first write, 
\begin{equation}
  v_{\mathrm{eq}} \, = \, \frac{ R_{\mathrm{eq}} }{t_{\mathrm{eq}} } \,= \, \frac{1}{t_{\mathrm{eq}}} \,
   \left(\frac{t_{\mathrm{eq}}^3- \Omega_{\mathrm{m}}}{1- \Omega_{\mathrm{m}}} \right)^{2/3} \, ,
   \label{veq}
\end{equation}
\noindent
and then search the solution of this equation, which  is
\begin {equation}
t_{\mathrm{eq}} \, = \, \left(\frac{v_{\mathrm{eq}}^{3/2} \, (1-\Omega_{\mathrm{m}})}{2}+ 
 \frac{\sqrt{v_{\mathrm{eq}}^3 \, (1-\Omega_{\mathrm{m}})^2
+4 \, \Omega_{\mathrm{m}}}}{2} \right)^{2/3} \, .
\label{teq}
\end{equation}
\noindent
Since, $   v_{\mathrm{eq}}$ is known and $\Omega_{\mathrm{m}}$ is chosen,
 the above equation gives the value of $t_{\mathrm{eq}} $ and 
then of $R_{\mathrm{eq}}$ by (\ref{veq}).
The redshift at the equality of 
the two densities is  obtained from expression (\ref{Rt}) by,
\begin{equation}
(1+z_{\mathrm{eq}})  \, t_{\mathrm{eq}} \, = \, \frac{\Omega_{\mathrm{m}} \,  \varrho_{\mathrm{c,0}}}
{K_0 \, \varrho_{\mathrm{\gamma,0}}} \, = \, 1/v_{\mathrm{eq}}\, .
\end{equation}
\noindent
Equations (\ref{CC}) and (\ref{c2eq}) provide the constants $C_{\mathrm{rel}}$ and $c_2$.
The numerical values of the various input parameters seen above are the following ones,
\begin{eqnarray}
H_0  \, =\,3.2408 \cdot 10^{-18} h  \quad [s^{-1}]\, \nonumber \\
\varrho_{\mathrm{rel,0}} \,= \,K_0  \frac{a \,T^4}{c^2} \,= \,4.6485 \cdot 10^{-34} \, K_0  \, \quad [g/cm^3] \, \nonumber \\
\varrho_{\mathrm{m,0}}\,= \,\Omega_{\mathrm{m}} \, \varrho_{\mathrm{c,0}} \, = 
1.8788 \cdot 10^{-29} h^2 \, \Omega_{\mathrm{m}} \, \quad [g/cm^3] \, \nonumber  \\
v_{\mathrm{eq}} = \, \frac{R_{\mathrm{eq}}}{t_{\mathrm{eq}}} \, = \, 2.4741 \cdot 10^{-5} 
\frac{K_0}{\Omega_{\mathrm{m}} \, h^2} \, .    
\label{values}
\end{eqnarray}
\noindent
Table 1 gives for different values of $\Omega_{\mathrm{m}}$  the values of the useful parameters  for the model description in 
the early phases according to the above equations  and constants.


\begin{table*}[h!]  
\vspace*{0mm} 
 \caption{ {\footnotesize{Values of parameters at the equilibrium point between matter and density in the scale invariant
 models with $k=0$ and  various $\Omega_{\mathrm{m}}$, as well as the constants  $C_{\mathrm{rel}}$  and  $c_2$.
 The value of $h=0.70$ is chosen here together with a  number of neutrinos $N_{\nu}=3$.
The time  $t_{\mathrm{eq}}$ is the equilibrium time and  $ t_{\mathrm{in}}$  the initial time, both in a timescale where $t_0 =1$.
The time  $\tau_{\mathrm{eq}}$ is the equilibrium time in years,
 in the scale  where the present age of the Universe is  13.8 Gyr.  
}}} 
\begin{center} 
\scriptsize
\begin{tabular}{ccccccccclc}
$\Omega_{\mathrm{m}}$  &  $v_{\mathrm{eq}}$  & $t_{\mathrm{eq}}$ & $R_{\mathrm{eq}}$ & $(1+z)_{\mathrm{eq}}$ 
&  $\rho_{\mathrm{eq}}$   & $T_{\mathrm{eq}}[K]$ & $C_{\mathrm{rel}}$ & $c_2$ & $ \quad \tau_{\mathrm{eq}} \quad \quad
\quad \quad \quad t_{\mathrm{in}}$\\
\hline
 &   &   &   \\ 
 0.50& 1.697892E-04 &0.7937009 & 1.347619E-04 &7420.5 & 1.492821E-18 & 1.802135E+04&1.358314E-03&5.850414E-02&2.76864E+04
 .7937007\\ 
 0.30& 2.829820E-04 &0.6694343 & 1.844378E-04 & 5278.8 & 2.719635E-19 & 1.177370E+04&6.930173E-04&5.874317E-02&5.66737E+04
 .6694333\\
 0.20& 4.244731E-04 &0.5848066 & 2.482347E-04 & 4028.4 & 7.039421E-20 & 8.397878E+03&5.305913E-04&6.735331E-02&1.01360E+05
 .5848043\\
 0.10& 8.489461E-04 &0.4641698 & 3.940551E-04 & 2537.7 & 6.983735E-21 & 4.713107E+03&4.192327E-04&9.503621E-02&2.80515E+05
 .4641616\\
0.04&  2.122365E-03 &0.3420487 & 7.259523E-04 & 1377.5 & 3.292346E-22 & 2.196149E+03&3.684662E-04&1.641062E-01&1.12216E+04
.3420085\\
\hline
\normalsize &
\end{tabular}
\end{center}
\end{table*}

\section{Physical Properties in the radiative era}   \label{physrad}

We now examine the evolution of the physical properties during the radiative era. The conservation laws of matter 
and radiation densities  (\ref{conserv}), together with the  constants given by (\ref{values}) lead to
\begin{eqnarray}
\varrho_{\mathrm{m}} (t) \,=  \,1.8788 \cdot 10^{-29} \, h^2 \, \Omega_{\mathrm{m}} \frac{t}{R^3(t)} \,  \quad [g/cm^3] \, , \nonumber \\
\varrho_{\mathrm{rel}} (t) = K_0 \, \varrho_{\gamma}\, =
 \, 4.6485  \cdot 10^{-34}\,  K_0 \, \frac{t^2}{R^4(t)} \,  \quad [g/cm^3] \, , \nonumber \\
T(t) \,= \, 2.726 \, \frac{t^{1/2}}{R(t)} \, \quad [°K] \,.
\label{temp}
\end{eqnarray}
\noindent
There, $R(t)$ is given by Equation (\ref{R}).
The initial  $t_{\mathrm{in}}$   of the Universe for a given model is obtained by setting $R(t)=0$ in this equation,
\begin{equation}
t_{\mathrm{in}} \, = \, \frac{C^{1/4}_{\mathrm {rel}}}{c^{1/2}_2} \,  ,
\label{tin}
\end{equation}
\noindent
which applies in the scales where 
$t_0=1$ and  $R_0=1$. We notice that the values obtained by this last expression 
are very close to those obtained by imposing $R(t)=0$ in Equation (\ref{jesus}) for the matter-dominated era,
which gives simply $t_{\mathrm{in}}= \Omega_{\mathrm{m}}^{1/3}$. The more appropriate values (\ref{tin}) for the initial time
in the radiative era
 are generally  slightly larger than $\Omega_{\mathrm{m}}^{1/3}$,  with a minor difference concerning only the sixth decimal.
  (It may also be recalled  \citep{Maeder17a} that $t_{\mathrm{in}}=0$ 
in the above mentioned scale only when $\Omega_{\mathrm{m}}=0$.)

If one considers very early epochs of the Universe, we may write  the time as $t = t_{\mathrm{in}}+ \Delta t $.
 This quantity $\Delta t$  represents  the age counted from the origin of the Universe, 
in a scale where the interval $1-t_{\mathrm{in}}$ is the present age of the Universe. 
Thus, the age  $\tau$, the cosmic time expressed in seconds, for a given
event at the time $t$ may be written,
\begin{equation}
\tau \, = \, 4.355 \cdot 10^{17}  \, \frac{\Delta t}{1- t_{\mathrm{in}}} \; [s] \,,
\label{tau}
\end{equation}
\noindent
where a present age of 13.8 Gyr has been adopted.
 This relates the ages $\tau$ and $\Delta t$ for a given cosmological model (since $t_{\mathrm{in}}$ depends
 on the model).  In the very early stages, the quantity $\Delta t$ is very small.
For example, for the cosmological nucleosynthesis which occurs during, say, the first three minutes $\Delta t < 4.1 \cdot 10^{-16}$,
 whatever the cosmological model.
Before the  equilibrium time of matter and radiation, say of the order of $10^5$ yr, one 
has $\Delta t < 7 \cdot 10^{-6}$.
Thus, we may consider that during the whole radiative phase  the 
value of $\Delta t$ is  $\ll 1$ for any flat scale invariant model.  This leads to an interesting  simplification of the expression of $R(t)$ 
given by Equation (\ref{R}). We may write,
\begin{equation}
t^4 \,=\, (t_{\mathrm{in}}+\Delta t)^4 \, \cong \, t^4_{\mathrm{in}}+ 4 \, t^3_{\mathrm{in}}\, \Delta t \, , \quad \mathrm{and}
 \quad t^2 \,  \, \cong   \,  t^2_{\mathrm{in}}+ 2 \, t_{\mathrm{in}}\, \Delta t \, .
 \end{equation} 
 \noindent
 The expansion factor becomes,
 \begin{equation}
 R(t) \, = 
 \, \frac{1}{\sqrt{2}} \left(c_2 \, t^4 -\frac{C_{\mathrm{rel}}}{c_2} \right)^{1/2} \,\cong \,  \frac{1}{\sqrt{2}} 
 \left(c_2 \,  t^4_{\mathrm{in}}+4\, c_2 \,  t^3_{\mathrm{in}} \Delta t -\frac{C_{\mathrm{rel}}}{ c_2 }\right)^{1/2} \, ,
 \end{equation}
 \noindent
 which may then be written  with  expression (\ref{tin}), 
 \begin{equation}
 R(t) \, \cong \, \sqrt{2} \, c^{1/2}_2 t_{\mathrm{in}}^{3/2}\, \Delta t^{1/2} \, ,
 \end{equation}
 \noindent
 and
 \begin{equation}
 R(t) \,  \cong\, B \, \Delta t^{1/2} \, ,
 \label{r}
 \end{equation}
 \begin{equation} 
  \mathrm{with} \; \;B = \sqrt{2} \, c^{1/2}_2 t_{\mathrm{in}}^{3/2}\, = \, \sqrt{2} \, \,\frac{C^{3/8}_{\mathrm{rel}}}{c^{1/4}_2} \, .
  \label{B}
  \end{equation}
 \noindent
 Thus, the time dependence of the expansion factor $R(t)$ for  the early Universe in the scale invariant context  is the same
 as in standard models. The Hubble term $H$ becomes
 \begin{equation}
 H(t) \, = \, \frac{1}{2} \, \frac{1}{\Delta t} \, ,
 \label{hdt}
 \end{equation}
 which applies to the scale where $t_0=1$. 
    \begin{figure}[t!]
\centering
\includegraphics[angle=0.0,height=8cm,width=13cm]{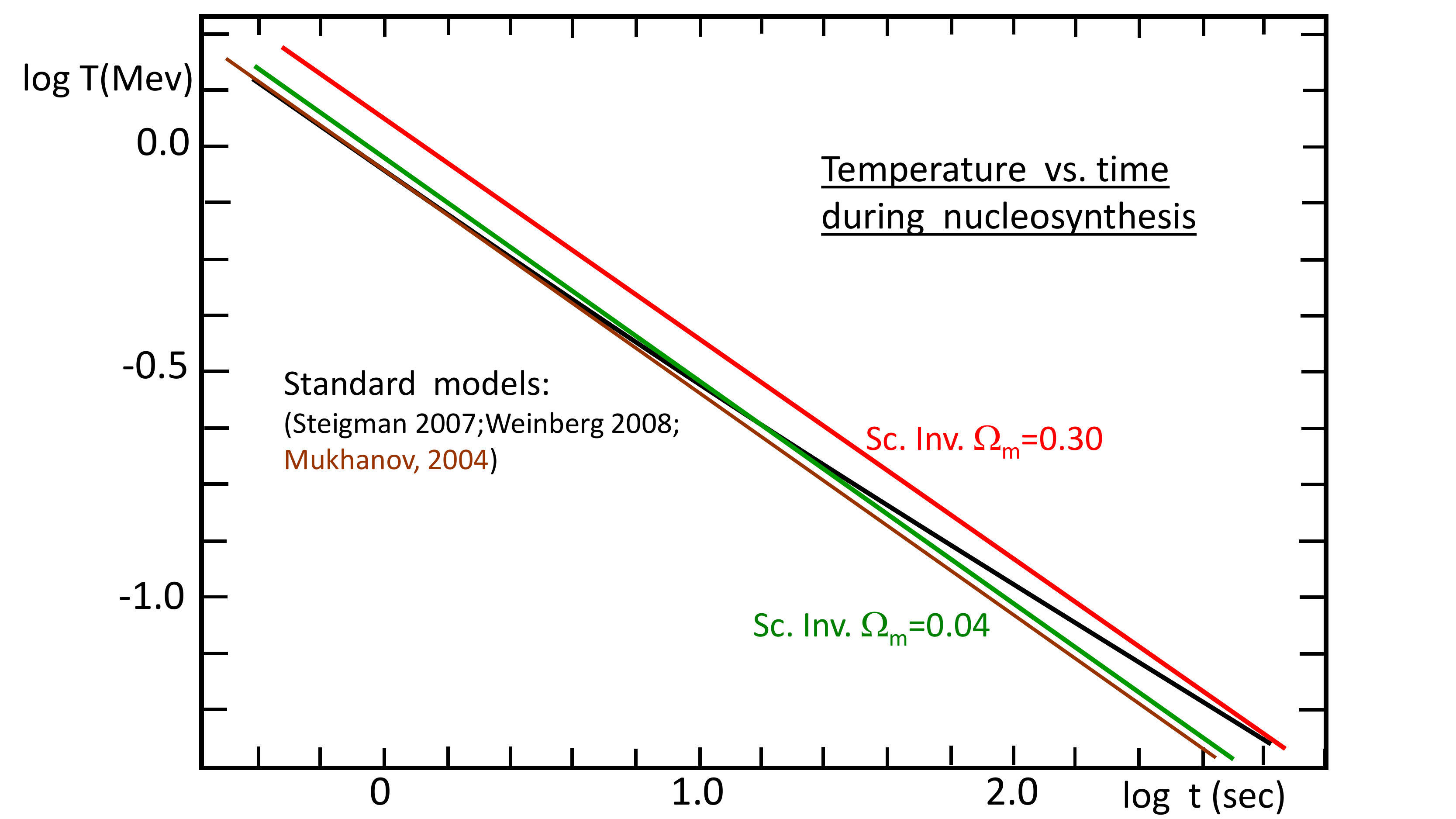}
\caption{Temperatures in Mev as a function of time in seconds for some models. The black line shows the values given
by \citet{Steigman07} and \citet{Weinberg08}, which are very similar. The brown line is the standard model based on
the value of the Planck temperature and Planck time scale as given by \citet{Mukhanov04}. The red  and green lines
correspond to   the scale invariant models for $\Omega_{\mathrm{m}}=0.30$  and $\Omega_{\mathrm{m}}=0.04$
respectively, with in both cases a value $h=0.7$ and a number of neutrino $N_{\nu}=3$. 
        }
\label{Temp}
\end{figure} 
 In the early phases, we can approximate expressions (\ref{temp}) by,
 \begin{eqnarray}
\varrho_{\mathrm{m}} (t) \, \cong \,  1.8788 \cdot 10^{-29}  h^2 
 \Omega_{\mathrm{m}} \frac{t_{\mathrm{in}}}{R^3(t)}  \, \cong \,    \frac{1.8788 \cdot 10^{-29}}{2^{3/2}}  h^2 
 \Omega_{\mathrm{m}} \frac{c^{1/4}_2}{C^{7/8}_{\mathrm{rel}}} \, \frac{1}{\Delta t ^{3/2}}
  \quad [g/cm^3] \, , \nonumber \\  
\varrho_{\mathrm{rel}} (t)  \, \cong \,  \, \cong \,  K_0 \, \varrho_{\gamma} =
  4.6485  \cdot 10^{-34}\,  K_0 \, \frac{t^2_{\mathrm{in}}}{R^4( t)}  \, \cong \,  
\frac{  4.6485  \cdot 10^{-34}}{4}\,  K_0  \, \frac{1}{C_{\mathrm{rel}}} \, \frac{1}{\Delta t ^2} 
  \quad [g/cm^3] \, , \nonumber \\  
T(t)  \, \cong \,  2.726 \, \frac{t^{1/2}_{\mathrm{in}}}{R( t)}  \, \cong \,  
\frac{2.726}{\sqrt{2}} \frac{1}{C^{1/4}_{\mathrm{rel}}} \, \frac{1}{\Delta t^{1/2} }
\, \quad [°K] \,.
\label{temp2}
\end{eqnarray}
\noindent
We have used the fact that $t= t_{\mathrm{in}} + \Delta  t  \, \cong \,  t_{\mathrm{in}}$.
The expansion factor
is given by Equation (\ref{r}), with Equations (\ref{tin}) and (\ref{B}) for 
$ t_{\mathrm{in}} $ and $B$. The values of $C_{\mathrm{rel}}$ and $c_2$ for different models
are given in Table 1. In the above expressions, the time $\Delta t$ is in the scale where $t_0 =1$.  We need these equations 
in the current units. With Equation (\ref{tau}), we have with  the time $\tau$ in seconds,
\begin{eqnarray}
\varrho_{\mathrm{m}} (\tau)  \, \cong \,       1.9091\cdot 10^{-3}\,  h^2 
 \Omega_{\mathrm{m}} \frac{c^{1/4}_2}{C^{7/8}_{\mathrm{rel}} (1-t_{\mathrm{in}})^{3/2}} \, \frac{1}{\tau^{3/2}}
  \quad [g/cm^3] \, , \nonumber \\  
\varrho_{\mathrm{rel}} (\tau)  \, \cong \,   22.0409 \, K_0  \, \frac{1}{C_{\mathrm{rel}}(1-t_{\mathrm{in}})^2} \, \frac{1}{\tau ^2} 
  \quad [g/cm^3] \, .
  \label{denstau}
  \end{eqnarray}
 \noindent
 The temperature expressed in  °K  or in MeV as a function of the time in sec is
 \begin{eqnarray}
 T(\tau)  \, \cong \,  \frac{1.272 \cdot 10^9}{C^{1/4}_{\mathrm{rel}} (1-t_{\mathrm{in}})^{1/2}} \, \frac{1}{\tau^{1/2}} \, 
 \quad [°K] \,, \nonumber \\
 T(\tau)   \, \cong \, \frac{0.1096}{C^{1/4}_{\mathrm{rel}} (1-t_{\mathrm{in}})^{1/2}} \, \frac{1}{\tau^{1/2}} \, \quad [MeV] \,.
 \label{Ttau}
 \end{eqnarray}
 \noindent
 The Hubble rate expressed in seconds is, consistently with Equations (\ref{tau}) and (\ref{hdt}),
 \begin{eqnarray}
 H(\tau) \ \cong \, \frac{1}{2} \, \frac{1}{\tau}  \quad  [s^{-1}]\, ,
 \label{Htau}
 \end{eqnarray}
 \noindent
 which is the same as in standard models.
 The above relations allow us to describe the physical conditions in the scale invariant  models with $k=0$ of the early Universe.
 Different values of the density parameter $\Omega _{\mathrm{m}}$ lead to different models.

 \section{The physical conditions at the time of the cosmological nucleosynthesis}

    \begin{figure}[t!]
\centering
\includegraphics[angle=0.0,height=9cm,width=18cm]{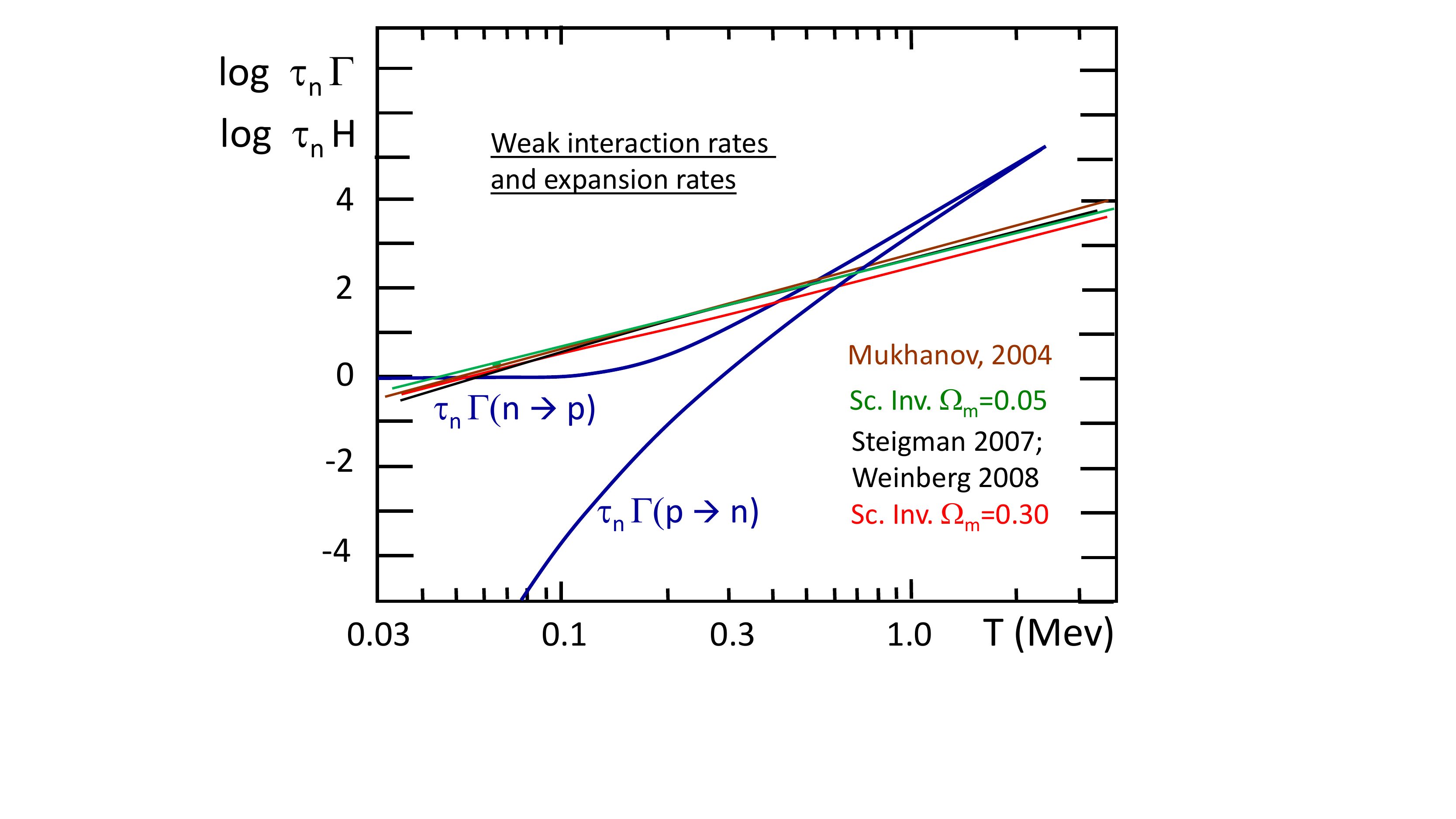}
\vspace{-10mm}
\caption{The weak interaction rates  $\Gamma (n \rightarrow p) \, \tau_{\mathrm{n}}$  and 
   $\Gamma (p \rightarrow n) \,\tau_{\mathrm{n}}$ (in sec$^{-1}$) 
   multiplied by the mean neutron lifetime $\tau_{\mathrm{n}}=885.7 s$
    as a function of  temperature  in MeV \citep{Durrer08}.  These rates are compared to the expansion rate $\tau_{\mathrm{n}} \, H$
    of different cosmological models as indicated. Same remarks as for Fig. 1.
        }
\label{rates}
\end{figure}

It is interesting to compare  the physical conditions predicted by the scale invariant models 
at the time of the nucleosynthesis with the corresponding values in standard models. The key parameters are the 
  temperature $T$  and the expansion rate $H$.
 The runs  $T(\tau)$, with the time $\tau$ is  in seconds, are illustrated for some standard and scale invariant
 models in Fig. \ref{Temp}. The  results of the standard models by  \citet{Steigman07} and \citet{Weinberg08} are
 very similar (black line), while the results by \cite{Mukhanov04} are slightly different. We see that the scale invariant models
with $\Omega_{\mathrm{m}}=0.30$ show  a significantly higher temperature at a given time than the standard models.
For  $\Omega_{\mathrm{m}}=0.04$, there is a nice agreement. This is interesting since in the scale invariant 
context, there is no need to call for an unknown dark matter, thus the matter content  $\Omega_{\mathrm{m}}$
 is only that resulting from the baryons, currently of the order of 0.04 \citep{Frie08}.

 At energies above 1 MeV, when electrons and positrons are still in equilibrium with radiation,
the equilibrium between neutrons and protons is maintained by the weak interactions,
$\nu + n  \leftrightarrow p+e^{-}, \, n+e^{+} \leftrightarrow p+\overline{\nu}$ and  $n \rightarrow p+e^{-}+\overline{\nu}.$
The ratio of the numbers of neutrons to protons is then defined by a Boltzman distribution of the form 
 $\frac{n_{\mathrm{n}}} {n_{\mathrm{p}}} \approx \, e^{Q/(kT)}$, 
 where $ Q= 1.293 $  MeV is the energy corresponding to the mass difference of neutrons and protons.
 At some  energy  below 1 MeV,  the  breakdown of the equilibrium  makes
  the first two  exchange rates to become rapidly negligible, only the third one is remaining, slowly reducing the number of free neutrons
  until deuterium is formed.
  
  Fig. \ref{rates}  shows as a a function of temperature  
  the rates  $\Gamma (n \rightarrow p) \, \tau_{\mathrm{n}}$  and 
   $\Gamma (p \rightarrow n) \,\tau_{\mathrm{n}}$ \citep{Durrer08}, 
   the $\Gamma$-rates are  the products  $\sigma \, v \, n$, expressed in sec$^{-1}$, of the corresponding
    cross-sections, velocities and concentrations.
   At energies above 1 MeV, we see  the  equality of the two rates.  Below it, the rate $\Gamma (p \rightarrow n)$
 rapidly vanishes,  while the neutron  slowly decays to protons at a timescale $\tau_{\mathrm{n}} = 885.7$ sec,
  so that  the product $\Gamma (n \rightarrow p) \tau_{\mathrm{n}}$ tends towards unity. 
  The expansion rate $\tau_{\mathrm{n}} \, H$
  is also illustrated.  When the $\Gamma$-rates are higher than the expansion rate $H$, the nuclear process are faster and thus  tend
  towards equilibrium. At the opposite, when the $\Gamma$-rates become lower, 
  the cosmological expansion  and the 
  associated  cooling rapidly dominate, the nuclear interactions cease, except for the neutron disintegration.
  The crossing temperature of the  $\tau_{\mathrm{n}} \, H$-  and $\tau_{\mathrm{n}} \,\Gamma $-lines defines the critical temperature,
  fixing the fraction of neutrons, which after the disintegration of a small part of them, 
will finally be turned to helium-4,  when deuterium formation starts 
at 0.085 MeV,

Fig. \ref{rates} also shows the expansion  rates    $\tau_{\mathrm{n}} \, H$  of the standard models 
together with the values of the scale invariant models already considered above. 
From Equations (\ref{Ttau}) and (\ref{Htau}), we  eliminate the time $\tau$  and obtain the relation
between $H$   in s$^{-1}$ and $T$ in MeV,
\begin{equation}
H(\tau) \, = \,41.6245 \,  \cdot C^{1/2}_{\mathrm{rel}} \,(1-t_{\mathrm{in}})  \,T^2 \,.
\end{equation}
\noindent
We see as before that
 the scale invariant model with $\Omega_{\mathrm{m}}=0.30$ somehow deviates from the standard models and would
 suggest a lower value of the freezing temperature for the neutron-proton decoupling. The model with a lower matter density
 (baryonic)  of $\Omega_{\mathrm{m}}=0.04$ shows an excellent agreement with the standard models,
the minor  differences being of the same order as the differences between the  various  standard models.  
The estimated  freezing  energy for neutrons  is around 0.6 to 0.7 MeV. The agreement of the
standard models and the low density scale invariant model is encouraging and makes worth  a detailed study 
of the cosmological nucleosynthesis in the scale invariant framework.

\section{Conclusions}

The runs of $T$(MeV) and of the expansion rate $H$ are essential parameters for the cosmological nucleosynthesis, 
which is used  in standard models to fix the   value of the density parameter $\Omega_{\mathrm{m}}$, typically found 
around 0.30. It is thus amazing that the scale invariant models, which usually do not need the presence of dark matter
as shown by a number of tests
\citep{Maeder17a,Maeder17b,Maeder17c, Maeder18,MGueorguiev18}, find an excellent agreement with the conditions
of the standard models for much lower values of $\Omega_{\mathrm{m}}$ of about 0.04. Such a density parameter leads
to the same value of the freezing energy of the neutron to proton ratio as in standard models.
These positive results are stimulating for the further study of the cosmological nucleosynthesis in the scale invariant context,
they come after a series of other positives tests as mentioned in the introduction.

 \vspace*{1cm} 
Acknowledgments: 
The author expresses his deep gratitude to his wife and to D. Gachet for their encouragements, as well to Dr. Vesselin G. Gueorguiev
for his continuous support and constructive remarks.




\vspace*{1cm}




\begin{thebibliography}{}


\bibitem[Bondi(1990)]{Bondi90} Bondi, H. 1990, in Modern Cosmology
in Retrospect, Eds. {Bertotti}, B., {Balbinot}, R., \& {Bergia}, S.,{Cambridge Univ. Press.}, 426 pp.

\bibitem[{{Bouvier} \& {Maeder}(1978)}]{BouvierM78}
{Bouvier}, P. \& {Maeder}, A. 1978, \apss, 54, 497

\bibitem[Bronstein \& Semendiaev(1974)]{Bronstein74} Bronstein, L.N., Semendiaev, K.A. 1974, Aide-memoire de mathematiques, 
Ed. Eyrolles, Paris, 935 p.

 
\bibitem[{{Canuto} {et~al.}(1977){Canuto}, {Adams}, {Hsieh}, \&
 {Tsiang}}]{Canu77}
{Canuto}, V., {Adams}, P.~J., {Hsieh}, S.-H., \& {Tsiang}, E. 1977, \prd, 16, 1643



\bibitem[{{Carroll} {et~al.}(1992){Carroll}, {Press}, \& {Turner}}]{Carr92}
{Carroll}, S.~M., {Press}, W.~H., \& {Turner}, E.~L. 1992, \araa, 30, 499

 

\bibitem[Carter(1979)]{Carter79} Carter, B. 1979, in ''Confrontation of cosmological theories with observational data'', 
IAU Symp. 63, Reidel Publ. Co., Dordrecht, p. 291.

\bibitem[Coles \& Lucchin(2002)]{Coles02} Coles, P., Lucchin, F. 2002, Cosmology. The Origin and Evolution of Cosmic Structure,
Wiley \& Sons Ltd, 492 p.


\bibitem[{{Dirac}(1973)}]{Dirac73}
{Dirac}, P.~A.~M. 1973, Proceedings of the Royal Society of London Series A,
 333, 403

\bibitem[Durrer(2008)]{Durrer08} Durrer, R. 2008, The Cosmic Microwave Background, Cambridge Univ. Press, 401 p.





 
\bibitem[{Frieman} { et~al.}(2008)]{Frie08}
{Frieman}, J.~A., {Turner}, M.~S., \& {Huterer}, D. 2008, \araa, 46, 38


\bibitem[Jesus(2018)]{Jesus18} Jesus, J.F. 2018, arXiv:1712.00697


\bibitem[Maeder(2017a)]{Maeder17a} Maeder, A. 2017a, \apj, 834, 194 

 \bibitem[Maeder(2017b)]{Maeder17b} Maeder, A. 2017b, \apj, 847, 65
 
 \bibitem[Maeder(2017c)]{Maeder17c} Maeder, A. 2017c, \apj, 849, 158
 
 \bibitem[Maeder(2018)]{Maeder18} Maeder, A. 2018, A. arXiv:1804.04484
 
 \bibitem[Maeder \& Bouvier (1979)]{MBouvier79} Maeder, A., Bouvier, P. 1979, Astron. Astrophys., 73, 82
 
 \bibitem[Maeder \& Gueorguiev(2018)]{MGueorguiev18}  Maeder, A., Gueorguiev, V. 2018, arXiv:1811.03495
 

\bibitem[Mukhanov(2004)]{Mukhanov04} Mukhanov, V. 2004, 
Intnl. J. Theoretical Physics, 43, 669

\bibitem[Steigman(2007)]{Steigman07} Steigman, G. 2007, Ann. Rev.
Nuclear and Particle Sci. 57, 463

\bibitem[Weinberg(2008)]{Weinberg08} Weinberg, S. 2008, Cosmology,
Oxford Univ. press, 593 p.

 
\bibitem[Weyl(1923)]{Weyl23}
{Weyl}, H. 1923, Raum, Zeit, Materie. Vorlesungen {\"u}ber allgemeine
 Relativit{\"a}tstheorie. Re-edited by Springer Verlag, Berlin, 1970




 
 \end{thebibliography}
\end{document}